\documentclass[fleqn,usenatbib]{rasti}

\usepackage{newtxtext,newtxmath}

\usepackage[T1]{fontenc}

\DeclareRobustCommand{\VAN}[3]{#2}
\let\VANthebibliography\thebibliography
\def\thebibliography{\DeclareRobustCommand{\VAN}[3]{##3}\VANthebibliography}

\usepackage{graphicx}           
\usepackage{amsmath}            
\usepackage{hyperref}           
\usepackage{xspace}             
\usepackage{tabularx}           
\usepackage{threeparttable}     

\newcommand{\eq}[1]{Equation~{#1}}
\newcommand{\fig}[1]{Figure~{#1}}
\newcommand{\hitops}{\texttt{HITOPS}\xspace}
\newcommand{\primes}{\texttt{PRIMES}\xspace}
\newcommand{\ffsc}{FF+SC\xspace}
\newcommand{\pwc}{PWC\xspace}

\title[Non-integer-$R$ FPT]{Frequency Phase Transfer for Future Millimetre Arrays with Arbitrary Frequency Ratios}

\author[S. S. Simelane et al.]{
Senkhosi S. Simelane,$^{1}$\thanks{E-mail: senkhxsi@gmail.com}
A. J. Kemball,$^{2}$
R. P. Deane$^{1,3,4}$
\\
$^{1}$Wits Centre for Astrophysics, School of Physics, University of the Witwatersrand, 1 Jan Smuts Avenue, Johannesburg 2001, South Africa\\
$^{2}$Department of Astronomy, University of Illinois Urbana-Champaign, 1002 W. Green St, Urbana 61801, United States of America\\
$^{3}$Department of Physics, University of Pretoria, Lynnwood Rd, Hatfield, Pretoria 0002, South Africa\\
$^{4}$Inter-University Institute for Data Intensive Astronomy, Department of Astronomy, University of Cape Town, Rondebosch, Cape Town  7701, South Africa
}

\date{Accepted XXX. Received YYY; in original form ZZZ}

\pubyear{\the\year{}}

\begin{document}
\label{firstpage}
\pagerange{\pageref{firstpage}--\pageref{lastpage}}
\maketitle

\begin{abstract}
Non-dispersive tropospheric turbulence-induced phase delays enforce significant, often dominant, limitations to the imaging fidelity and dynamic range in (sub-)millimetre astronomy. Frequency Phase Transfer (FPT), which removes such delays from high-frequency data using simultaneous lower-frequency observations, has become increasingly viable with the advent of shared-optical-path multi-band receivers and is a key motivator of the Event Horizon Telescope Collaboration’s ambitions to add 86-GHz and 345-GHz bands alongside its existing 230-GHz band. However, existing FPT algorithms break down for non-integer frequency ratios, leaving jump discontinuities in the residual phases. We introduce a new FPT algorithm, phase-wrap counting (\pwc), which works for any frequency ratio and clarifies the nature and source of the jump discontinuities left by previous FPT approaches. Using the newly developed High-frequency Inter-band Transfer of Phase Solutions (\hitops) software package, we apply \pwc to a simulated dual-band observation with the Event Horizon Telescope and demonstrate its effectiveness under realistic atmospheric conditions and thermal noise levels. The method successfully calibrates the 230-GHz phases using the 86-GHz phase solutions, outperforming the conventional phase calibration approach in both coherence and peak flux recovery. This result indicates that \pwc has the potential to enable the calibration of the 230-GHz band using the 86-GHz band in future EHT operations.
\end{abstract}




\section{Introduction}

 Low source brightness and comparatively high system temperatures hinder the calibration of high-frequency very-long-baseline interferometry (VLBI) observations. The elevated system temperatures at (sub-)millimetre frequencies arise from increased atmospheric opacity, and thus increased atmospheric self-emissivity. Longer integration times can alleviate these issues, but this approach is infeasible without adequate correction of rapid phase fluctuations induced by tropospheric turbulence. The idea of using simultaneous or near-simultaneous observations at a lower reference frequency to stabilise faster-varying phases at a higher target frequency, by extrapolating the non-dispersive reference tropospheric phase delays, has been explored for a few decades. Early such experiments include \citet{Asaki1998}, \citet{Middelberg2005}, and \citet{Rioja2005}. However, the instrumental setups used had unavoidable limitations related to slightly different receiver optical paths \citep{Asaki1998} and the requirement of temporal interpolation due to fast frequency-switching rather than simultaneous multi-frequency observations \citep{Middelberg2005, Rioja2005}. Despite the experimental limitations, these early works demonstrated the promise and feasibility of such approaches.\par

\citet{Dodson2008, Dodson2009} introduced the term \emph{Frequency Phase Transfer} (FPT) to describe the use of concurrent observations of the same source at different frequencies to mitigate tropospheric turbulence-induced phase corruptions. FPT extends the effective coherence time of the atmosphere, enabling the use of longer time-averaging intervals during data calibration to achieve higher signal-to-noise ratios (SNRs), visibility detection rates, and image fidelity.\par

Thanks to significant breakthroughs in shared-optical-path receiver technology at the Korean VLBI Network \citep[KVN;][]{Han2008, Han2013}, simultaneous multi-frequency observations are now possible \citep[see][for example]{Dodson2014, Rioja2014, Algaba2015, Yoon2018, Zhao2019}. Through FPT, these works have managed to extend coherence times from tens of seconds to a few tens of minutes at frequencies between 43\,GHz and 130\,GHz. FPT approaches have also been supplemented with steps to remove residual dispersive delays. These include source-frequency phase referencing \citep[SPFR;][]{Rioja2011}, multi-frequency phase referencing \citep[MFPR;][]{Dodson2017}, and FPT-squared \citep{Zhao2017}. These methods have led to the successful extension of coherence times to several hours, even at millimetre wavelengths \citep[130\,GHz;][]{Algaba2015, Rioja2015}.\par

The well-demonstrated efficacy of FPT has brought it keenly into the focus of even higher-frequency VLBI projects, such as the next-generation Event Horizon Telescope \citep[ngEHT;][]{Blackburn2019, Johnson2023, Doeleman2023}. Owing to the success of current multi-frequency receiver technology and FPT techniques up to 130\,GHz, plans for the Event Horizon Telescope (EHT) and ngEHT to adopt a tri-band receiver system are underway, with the goal of enabling the leading-edge black hole shadow and jet physics science applications detailed by \citet{Jiang2022}, \citet{Issaoun2023}, and \citet{Johnson2023}, among others. The planned receiver system will operate simultaneously at 86, 230, and 345\,GHz \citep{Doeleman2023, EHT2024}, allowing significantly improved calibration of tropospheric phase fluctuations through FPT. In the longer term, FPT will also aid in the calibration of space-based extensions to the array \citep[e.g., the Black Hole Explorer;][]{Johnson2024}, which will introduce additional non-dispersive phase errors due to orbital position errors in the correlator model \citep{Rioja2011b}.\par

Recently, EHT stations have been used for single-baseline FPT test experiments by \citet{Issaoun2025} (86\,GHz to 215\,GHz) and \citet{Zhao2025} (86\,GHz to 258\,GHz). In both experiments, clear correlation was observed between the phases in the reference and target bands, and significant coherence improvements were found in the FPT residuals. This demonstrated the feasibility of the paired-antenna approach in the case of \citet{Issaoun2025}, and the viability of the current simultaneous multi-band receiver design in the case of \citet{Zhao2025}. While early FPT tests at these frequencies have been successful, a major problem with using the planned ngEHT bands to enable FPT calibration of future EHT data is yet to be addressed: the reliance of existing FPT methods on integer ratios between the reference and target frequencies. At non-integer frequency ratios, when wraps occur in the reference-band phases, the FPT residuals show jump discontinuities that have appeared to be random \citep{Rioja2023, Issaoun2025}. This currently prohibits straightforward 86-to-230\,GHz FPT calibration.\par

Solutions to this problem have been proposed by \citet{Dodson2014} and \citet{Rioja2023}. The former proposed using the nearest integer frequency ratio to extrapolate the fringe-fitting phase solutions in the reference band to a frequency just outside the target band. However, this requires high-quality delay calibration in the target band, and extrapolating to a frequency no further than one bandwidth from the edge of the band \citep{Dodson2014}. Even with the aspirational 16-GHz ngEHT bandwidth, this approach will not be viable. \citet{Rioja2023} proposed increasing the bandwidth of the planned 86\,GHz band to reach 115\,GHz, a demanding technical requirement, as this would require an increase of the planned bandwidth by a factor of 2.5. Moreover, pressure broadening of the 118.75-GHz molecular oxygen spectral line would likely hinder this solution \citep{Rosenkranz}. Since existing and upcoming EHT antennas will be equipped with 86- and 230-GHz bands before full tri-band deployment in the expanded array, it is prudent to explore possible alternatives to these two solutions.\par

In this paper, we explain the origin of the phase jumps seen in previous works \citep[e.g.,][]{Dodson2014, Rioja2023}, present a new FPT algorithm for correcting them, introduce its implementation in the High-frequency Inter-band Transfer of Phase Solutions (\hitops) software package, and compare its performance to the classical FPT algorithm using simulated dual-band data. We introduce the new FPT algorithm and software in Section~\ref{sect:methods}. There, we also describe the interferometric simulations used to generate the results presented and discussed in Section~\ref{sect:results_and_discussion}. Finally, we provide a summary of our findings in Section~\ref{sect:conclusions}, as well as plans and recommendations for further testing and validation.

\section{Methods} \label{sect:methods}

Radio signals of astrophysical origin encounter several corrupting effects between the source and the detector. Interferometric visibility phases at the reference frequency, $\nu_\mathrm{r}$, after application of the correlator model can be expressed as the sum of the contributions of the source and the residual corruptions \citep[following][]{Rioja2005}:
\begin{equation}
    \phi(t) = \phi_{\mathrm{src}}(t) + \phi_{\mathrm{geo}}(t) + \phi_{\mathrm{inst}}(t) + \phi_{\mathrm{ion}}(t) + \phi_{\mathrm{trop}}(t) + \phi_{\mathrm{ther}}(t) + 2\pi n(t)\,, 
\end{equation}
where $t$ is a time coordinate with origin $t_0$, the start of the recording of visibilities. Further, terms include those for source structure $\phi_{\mathrm{src}}(t)$, geometric delay $\phi_{\mathrm{geo}}(t)$, instrumental phase corruption $\phi_{\mathrm{inst}}(t)$, ionospheric propagation $\phi_{\mathrm{ion}}(t)$, tropospheric propagation $\phi_{\mathrm{trop}}(t)$, and thermal noise $\phi_{\mathrm{ther}}(t)$. The final term, $2\pi n(t)$, represents the inherent $2\pi$ phase ambiguity, with
\begin{equation}
    n(t) \in \mathbb{Z} \quad \forall \ t\,.
\end{equation}
We adopt a notation where lowercase symbols are used for terms at the reference frequency, and uppercase symbols for those at the target frequency, $\nu_{\mathrm{t}}$. The phase, $\Phi(t)$, recorded at $\nu_{\mathrm{t}}$ can then be expressed analogously as
\begin{align}
    \Phi(t) &= \Phi_{\mathrm{src}}(t) + \Phi_{\mathrm{geo}}(t) + \Phi_{\mathrm{inst}}(t) + \Phi_{\mathrm{ion}}(t) + \Phi_{\mathrm{trop}}(t) + \Phi_{\mathrm{ther}}(t) \nonumber \\
    &\quad + 2\pi N(t)\,. 
\end{align}    
Ionospheric delays are negligible at higher frequencies \citep[$\gg10$\,GHz;][]{Thompson2017}. Thus, we ignore $\phi_{\mathrm{ion}}(t)$ in this work. For simplicity, we also assume that the thermal noise term is negligible in comparison to the tropospheric terms, and temporarily omit it from consideration.

\subsection{Frequency Phase Transfer} \label{sect:fpt}
FPT seeks to leverage the non-dispersive nature of the tropospheric phase delays \citep{Thompson2017}:
\begin{equation}
    \Phi_{\mathrm{trop}}(t) = R \phi_{\mathrm{trop}}(t)\,,
    \label{eqn:nondispersive_delays}
\end{equation}
where, following common notation, $R \equiv \nu_\mathrm{t} / \nu_{\mathrm{r}}$. Fringe-fitting and self-calibration at $\nu_{\mathrm{r}}$ can solve for and separate $\phi_{\mathrm{src}}(t)$ from
\begin{equation}
    \phi_\mathrm{cal}(t) = \phi_\mathrm{geo}(t) + \phi_\mathrm{inst}(t) + \phi_\mathrm{trop}(t) + 2\pi n(t)\,.
    \label{eqn:phi_cal}
\end{equation}
Scaling the resultant $\phi_\mathrm{cal}(t)$ by $R$ and subtracting the product from $\Phi(t)$ gives \citep{Rioja2005, Dodson2009}
\begin{align}
    \Phi_{\mathrm{FPT}}(t) &= \Phi(t) - R \phi_\mathrm{cal}(t) \label{eqn:cfpt} \\
    &= \Phi_{\mathrm{src}}(t) + \left[ \Phi_{\mathrm{geo}}(t) - R \phi_{\mathrm{geo}}(t) \right] + \left[ \Phi_{\mathrm{inst}}(t) - R \phi_{\mathrm{inst}}(t) \right] \nonumber \\
    &\quad + \left[ \Phi_{\mathrm{trop}}(t) - R \phi_{\mathrm{trop}}(t) \right] + 2\pi  \left[N(t) - Rn(t) \right]\,. 
\end{align}
The second term on the right-hand side is equal to $2\pi \vec{D_{\lambda_t}(t)} \cdot \vec{\theta}(t)$, where $\vec{D_{\lambda_t}(t)}$ is the projected baseline vector in units of the wavelength $c / \nu_{\mathrm{t}}$ and $\vec{\theta}(t)$ the relative core shift between the frequencies \citep{Lobanov1998}. By \eq{\ref{eqn:nondispersive_delays}}, the third term in square brackets is equal to zero, and the instrumental term can be removed through fringe-fitting \citep[e.g.,][]{Algaba2015}, self-calibration, or phase referencing using a calibrator source \citep[e.g.,][]{Rioja2005, Rioja2011}. After this step,
\begin{equation}
    \Phi_{\mathrm{FPT}}(t) = \Phi_{\mathrm{src}}(t) + 2\pi \left[ \vec{D_{\lambda_t}(t)} \cdot \vec{\theta}(t) \right] + 2\pi  \left[ N(t) - Rn(t) \right] \,.
    \label{eqn:fpt_residuals}
\end{equation}
Since $N(t) - Rn(t)$ is an integer if the frequency ratio is an integer, this effectively leaves only the scientifically interesting phases,
\begin{equation}
    \Phi_0 (t) \equiv \Phi_{\mathrm{src}}(t) + 2\pi \left[ \vec{D_{\lambda_t}(t)} \cdot \vec{\theta}(t) \right]\,,
    \label{eqn:Phi_0}
\end{equation}
encoding information about the core shift and morphology of the source. This is the mathematical essence of FPT.

\subsection{The origin of jump discontinuities in FPT residuals}
It is evident from \eq{\ref{eqn:fpt_residuals}} that a problem arises when $R$ takes on a non-integer value because the final term on the right-hand side, $ 2\pi \left[ N(t) - Rn(t) \right]$, is no longer an immaterial integral $2\pi$ phase ambiguity. It leaves an observable contribution in the FPT residuals, manifesting as jump discontinuities in the residuals \citep[e.g.,][]{Dodson2014, Rioja2023, Issaoun2025}. However, the nature of this contribution has not always been fully understood.\par

To illuminate the origin of the jump discontinuities, we can write $R$ as the sum of its floor, $r$, and residual excess, $\epsilon$:
\begin{equation}
    R = r + \epsilon \,. 
\end{equation}
The FPT residuals can then be rewritten as
\begin{align}
    \Phi_{\mathrm{FPT}}(t) &= \Phi_0 (t) + 2\pi  \left[ N(t) - (r+\epsilon)n(t) \right] \nonumber \\
    &= \Phi_0 (t) + 2\pi \left[ N(t) - rn(t) \right] - 2\pi \epsilon n(t)\,. 
\end{align}
The quantity $N(t) - rn(t)$ is an integer, so the culprit of the issues encountered in non-integer-$R$ FPT can only be the final term,
\begin{equation}
    \delta \Phi_{\mathrm{FPT}}(t) \equiv - 2\pi \epsilon n(t)\,. 
\end{equation}
Recall that $n(t)$ is the integer associated with the ambiguity term in $\phi(t)$. Thus, $\delta \Phi_{\mathrm{FPT}}(t)$ is a constant phase offset that changes value abruptly when a phase-wrap occurs in $\phi(t)$, i.e., when $n(t)$ changes. This is the source of the jump discontinuities.\par

We can validate this result by checking if its predictions are consistent with offsets reported in other works. In Section 3.1 of \citet{Middelberg2005}, the authors present a hypothetical case in which the frequency ratio is $R=2.5$, and $\phi(t)$ wraps from $\phi(t_i)=359\,\mathrm{deg}$ to $\phi(t_{i+1})=0\,\mathrm{deg}$. The authors state that the scaled reference-frequency phases to be used for calibration will see a consequent jump from $R\phi(t_i)=897.5\,\mathrm{deg}$ ($=177.5\,\mathrm{deg}$) to $R\phi(t_{i+1})=0\,\mathrm{deg}$. A phase change of $1\,\mathrm{deg}$ at $\nu_\mathrm{r}$ corresponds to a change of $2.5\,\mathrm{deg}$ at $\nu_\mathrm{t}$. Therefore, the scaled reference-frequency phases undergo a jump of $177.5\,\mathrm{deg} + 2.5\,\mathrm{deg}= 180\,\mathrm{deg}$. For $R=2.5$,
\begin{equation}
    \epsilon = R - r = \frac{5}{2} -2 = \frac{1}{2}\,. 
\end{equation}
Assuming that $n(t)$ goes from $n(t_i)=0$ to $\left|n(t_{i+1})\right|=1$,
\begin{equation}
    \left|\delta \Phi_{\mathrm{FPT}}(t) \right| = \left| 2\pi \times \frac{1}{2} \right| = \pi\,, 
\end{equation}
matching their result.\par

Referring to Figure 5 in their paper, \citet{Rioja2023} noted that, for their simulations with $\nu_\mathrm{r} = 255\,$GHz and $\nu_\mathrm{t}=340\,$GHz, the phases changed by $\pm 120\,\mathrm{deg}$ at each phase-wrap in $\phi(t)$. For this frequency ratio,
\begin{equation}
    \epsilon = \frac{4}{3}-1 = \frac{1}{3}\,, 
\end{equation}
which yields
\begin{align}
    |\delta \Phi_{\mathrm{FPT}}(t)| &= \left| 2\pi \times \frac{1}{3} \right|  \\
    &= \frac{2}{3}\pi \,
\end{align}
when $|n(t)|=1$. Therefore,
\begin{equation}
    n(t) \in \{-1, 0, 1\} \quad \forall \ t 
\end{equation}
in their observation explains their results. This shows that the phase jumps are, in fact, deterministic. Given any integer ratio, we can predict the values of all phase offsets that could be introduced by non-integer-$R$ FPT. This ability empowers us to excise them from the residuals, if we can reliably track $n(t)$, by subtracting $\delta \Phi_{\mathrm{FPT}}(t)$ from the FPT residuals. We now introduce a new algorithm, \emph{phase-wrap counting} (\pwc), that seeks to exploit this fact.

\newcolumntype{L}[1]{>{\hsize=#1\hsize\arraybackslash}X}
\newcolumntype{C}[1]{>{\hsize=#1\hsize\centering\arraybackslash}X}
\newcolumntype{R}[1]{>{\hsize=#1\hsize\raggedright\arraybackslash}X}
\begin{table*}
    \centering
    \caption{Dual-band EHT array used in interferometric simulation.}
    \label{tab:eht_dualband_array}
    \begin{threeparttable}
    \begin{tabularx}{\linewidth}{L{1.25} C{0.75} C{0.75} C{0.75} C{0.75} R{1.75}}
        \hline
        Station & Code & Diameter (m) & \multicolumn{2}{c}{Receiver Temperature (K)} & Reference(s) \\
        & & & 86\,GHz & 230\,GHz & \\
        \hline
        APEX\tnote{a}   & AP & 12 & 40 & 85 & APEX website\tnote{b}\\
        SMA    & SM & 15 & 70\tnote{c} & 70 & \citet{Wilner1998}\\
        IRAM   & PV & 30 & 30 & 60 & \citet{Carter2012}\\
        GLT    & GL & 12 & 40 & 70 & \citet{Hasegawa2017}\\
        NOEMA  & PB & 50 & 50 & 80 & \citet{Chenu2016}\\
        KP     & KP & 12 & 40 & 80\tnote{d} & \citet{Pesce2024} \\
        KVNPC  & KC & 21 & 60 & 60 & \citet{Han2013}; \citet{Shin2024}\\
        KVNYS  & KY & 21 & 60 & 60\tnote{e} & \citet{Han2013}\\
        \hline
    \end{tabularx}
    \begin{tablenotes}
        \item[a] The APEX, GLT, and KP 86-GHz receiver temperatures are from ALMA Band 3 \citep{Claude2008}.
        \item[b] See \url{https://www.apex-telescope.org/ns/observing/the-telescope/instruments/nflash/}.
        \item[c] This is the receiver temperature of JCMT \citep{Han2018}, which observes at the reference frequency in the JCMT-SMA paired antenna.
        \item[d] Assumed to match SMT receiver temperature estimated by \citet{Pesce2024} from system temperature stated on the SMT website (see \url{https://aro.as.arizona.edu/?q=facilities/uarizona-aro-submillimeter-telescope}).
        \item[e] Assumed to match KNVPC receiver temperature reported by \citet{Shin2024}.
    \end{tablenotes}
    \end{threeparttable}
\end{table*}

\subsection{Phase-wrap counting}
To correct for the phase jumps introduced by FPT with non-integer $R$ given simultaneously recorded phases in two bands, we propose an algorithm that
\begin{enumerate}
    \item calculates $\epsilon$ as $\epsilon=R-r$;
    \item estimates $n(t)$ by assuming $n(t_0) = 0$ and tracking the number of phase-wraps that have occurred at each time stamp; and
    \item computes the FPT-calibrated phases,
        \begin{align}
            \Phi_{\mathrm{PWC}}(t) &= \Phi_{\mathrm{FPT}}(t) - \delta \Phi_{\mathrm{FPT}}(t) \nonumber \\
            &= \Phi(t) - R \phi_\mathrm{cal}(t) + 2\pi \epsilon n(t) \nonumber \\
            &= \Phi_0 (t) + 2\pi \left[ N(t) - rn(t) \right]\,. 
            \label{eqn:pwc}
        \end{align}
\end{enumerate}
This procedure, in effect, recovers the phases dominated by source structure and geometric information (\eq{\ref{eqn:Phi_0}}), 
\begin{equation}
    \Phi_{\mathrm{PWC}}(t) = \Phi_0 (t)\,,
    \label{eqn:pwc_residuals}
\end{equation}
assuming that the estimate of $n(t)$ obtained in step (ii) is correct. Henceforth, we shall refer to the conventional FPT calibration procedure represented by \eq{\ref{eqn:cfpt}}, simply entailing the subtraction of $R\phi_\mathrm{cal}(t)$ from $\Phi(t)$, as \emph{classical FPT} (CFPT). CFPT has been used successfully in several previous works for FPT calibration of integer-$R$ data \citep[e.g.,][]{Rioja2005, Algaba2015, Rioja2015, Zhao2025}. In the non-integer-$R$ case, FPT calibration can, in principle, also be achieved by unwrapping $\Phi(t)$ and $\phi_\mathrm{cal}(t)$ before applying CFPT and then wrapping the resultant $\Phi_\mathrm{FPT}(t)$. We shall refer to this approach as \emph{classical FPT with unwrapping} (CWU), which, to our knowledge, has not been used in any published work. The term \emph{FPT} will henceforth refer to the class of non-dispersive delay correction algorithms that use inter-band phase transfer, including CFPT, CWU, and \pwc.\par

Having demonstrated that, in theory, non-integer-$R$ FPT is possible, some practical questions arise. Among the most important are how to track $n(t)$ and whether it can be tracked accurately in the presence of random noise. The former is simpler to address. When a wrapped phase time-series
\begin{equation}
    \phi_\mathrm{cal}(t) \in [-\pi,\pi) \quad \forall\ t\,
\end{equation}
crosses the phase boundary from the $-\pi$ side, it undergoes a phase jump of $+2\pi$. Conversely, a crossing from the $+\pi$ side produces a phase jump of $-2\pi$. Therefore, $n(t)$ increases by one when the phase boundary is crossed from the $-\pi$ side, and decreases by one when the boundary is crossed from the $+\pi$ side. We increment/decrement $n(t)$ whenever a phase jump of magnitude $>\pi$ is detected between consecutive samples of $\phi_\mathrm{cal}(t)$. To answer the question concerning the tracking of $n(t)$ in the presence of noise, we rearrange \eq{\ref{eqn:phi_cal}} to isolate $n(t)$:
\begin{equation}
    n(t) = \frac{\phi_\mathrm{cal}(t) - \left[ \phi_\mathrm{inst}(t) + \phi_\mathrm{trop}(t) + \phi_\mathrm{ther}(t) \right]}{2\pi}\,.
\end{equation}
This equation makes it clear that $n(t)$ depends on the \emph{sum} of the terms that make up $\phi_\mathrm{cal}(t)$, including thermal noise. That is, the noise is already embedded in the signal being tracked, rather than acting as a separate perturbation on $n(t)$. Thus, noise does not make the problem any harder, except in the extremely low-SNR regime, where noise can drive the true, unwrapped phase $\tilde{\phi}(t)$ to violate
\begin{equation}
    \left| \tilde{\phi} (t+\Delta t) - \tilde{\phi} (t) \right| < \pi \quad \forall t\,, 
\end{equation}
for integrations of duration $\Delta t$. This condition, known as the Itoh condition \citep{Itoh1982}, is necessary for exact phase tracking, as it ensures that all phase jumps
\begin{equation}
    \left| \phi_\mathrm{cal} (t+\Delta t) - \phi_\mathrm{cal} (t) \right| \geq \pi    
\end{equation}
will be legitimate phase-wraps. The requirement that the Itoh condition holds is a limitation \pwc shares with CWU, which other works \citep[e.g.,][]{Rioja2023, Issaoun2025} have noted becomes increasingly challenging in the low-SNR regime. However, \pwc has a key advantage over CWU: in \pwc, phase tracking is only performed on the reference-frequency phases, while CWU involves unwrapping both the reference- and target-frequency phases before the transfer of phase solutions. Thus, \pwc only requires that the Itoh condition holds at the reference frequency, whereas CWU requires that it holds at both frequencies. \pwc is therefore significantly less prone to the low-SNR problem because the SNR is generally much higher at the reference frequency. The advantage also makes \pwc less computationally expensive, but the extent to which this is beneficial depends strongly on the complexity of the unwrapping algorithm used.

\subsection{\hitops: A software package for FPT calibration}
Our \pwc algorithm is implemented within the new High-frequency Inter-band Transfer of Phase Solutions (\hitops) software package. Presented here for the first time, \hitops has been developed with scalability to large data volumes in mind. It employs \texttt{dask-ms} \citep{Perkins2024, Perkins2025} to read Measurement Set version 2 (MSv2; \citealt{Kemball2000}) data, and uses \texttt{dask-ms} with \texttt{dask} \citep{Rocklin2015} for parallel writing operations. The package therefore expects the reference- and target-frequency data in separate MSv2-format datasets, each averaged over frequency. The reference-frequency source structure can be optionally removed from the column to be used for FPT by specifying a model column for the reference-frequency data. By reshaping the visibility data to a form consistent with the forthcoming Measurement Set version 4 standard\footnote{See \url{https://xradio.readthedocs.io/en/stable/measurement_set/overview.html}.}, which facilitates efficient access to per-baseline data, calibration is parallelised across baselines using \texttt{dask}. The underlying functions are accelerated through just-in-time (JIT) compilation using \texttt{Numba}\footnote{See \url{https://numba.pydata.org/}.}, with FPT corrections applied per-correlation phase time-series. These corrections are applied according to the user’s selected algorithm, chosen from \pwc, CFPT, and CWU.\par

In addition to the \texttt{calibrate} mode, which performs the corrections described above, a \texttt{diagnose} mode provides functionality for evaluating and visualising FPT performance. In this mode, \hitops can generate diagnostic summaries in the form of plots and text files. The text files are designed to facilitate the identification of baselines or antennas where FPT calibration performed poorly, and include tables comparing a chosen coherence metric before and after FPT for each baseline, along with the percentage improvement of the metric. Baselines are ranked from worst to best by percentage improvement, and array-wide summary statistics such as the mean and median percentage improvements are also provided. The available coherence metrics include the coherence factor \citep{Thompson2017, Zhao2025},
\begin{equation}
    C(T) = \left | \frac{1}{T} \int_0^T e^{i\phi(t)}dt \right |\,,
    \label{eqn:coherence_factor}
\end{equation}
and the effective coherence time, $\tau_\mathrm{c}$, which we define as the expectation value of the time-dependent averaging interval that produces an amplitude loss of $e^{-1}$. When $C(T)$ is used as the performance metric, the averaging interval, $T$, can be specified by the user. Diagnostic plots include per-baseline comparisons of phases or of either coherence metric before and after FPT, as well as $(u,v)$-coverage maps coloured by the ratio of post- to pre-FPT coherence factors.

\begin{figure}
    \centering
    \includegraphics[width=\linewidth]{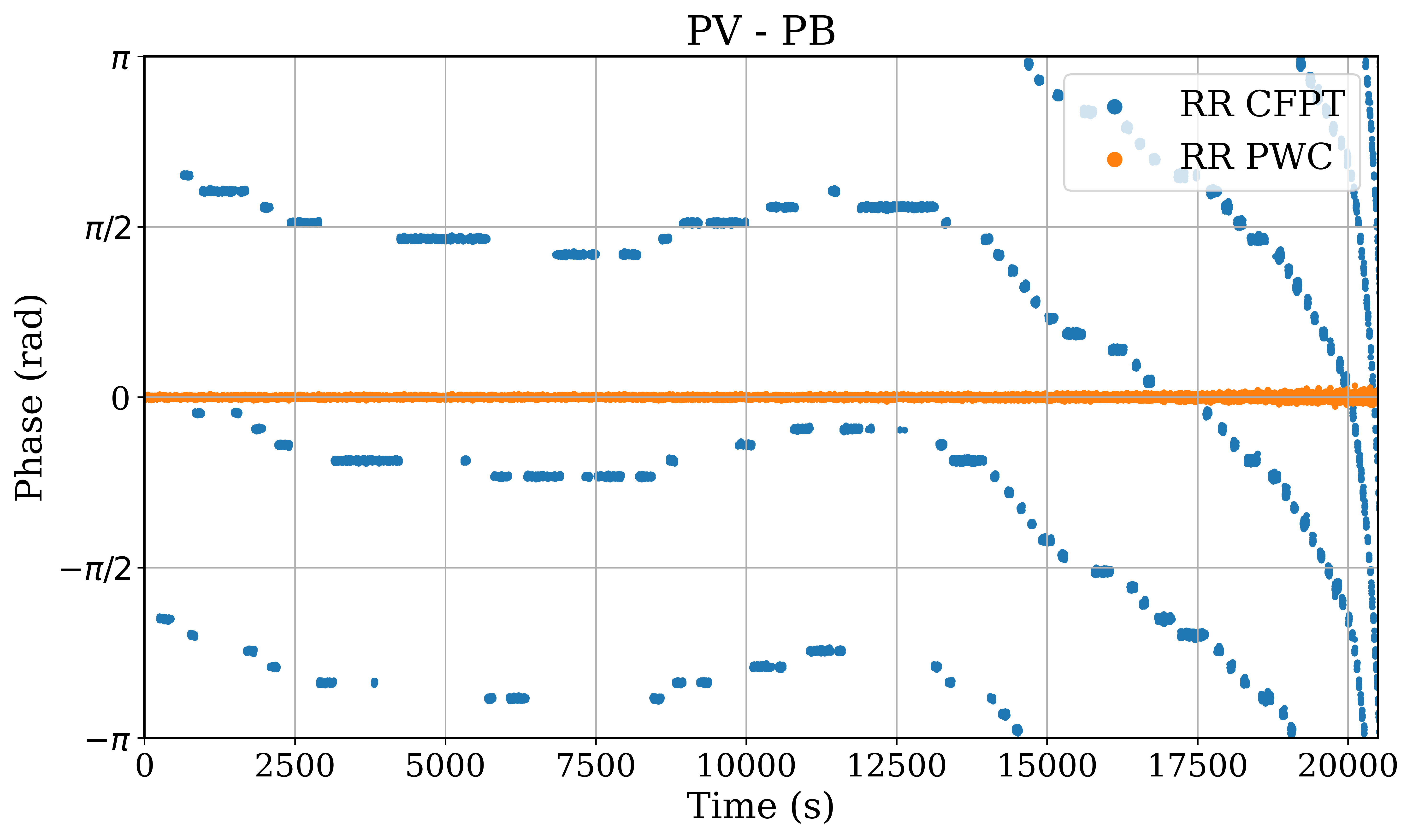}
    \caption{Residual target-frequency (230\,GHz) phase after CFPT (blue) and \pwc (orange) corrections for the baseline formed by IRAM and NOEMA. The CFPT residuals show jump discontinuities, while \pwc successfully stabilises the phases despite the non-integer frequency ratio, removing the phase jumps seen in the CFPT residuals.}
    \label{fig:cfpt_vs_pwc_phase}
\end{figure}

\subsection{Interferometric Simulations}
As a proof-of-concept demonstration, we applied \pwc to synthetic dual-band EHT data generated with the Parallelised Radio Interferometer Measurement Equation Simulator (\primes), the conceptual successor to \texttt{MeqSilhouette} \citep{Blecher2017, Natarajan2022}, which is also presented here for the first time. While retaining many of the core algorithms and scientific objectives of \texttt{MeqSilhouette}, \primes introduces an entirely new codebase optimised through parallelism and JIT compilation, and designed for future extensibility. Its primary goal is to extend the functionality of \texttt{MeqSilhouette} by enabling fast simulation of simultaneous multi-band observations and seamless incorporation of real weather data, which is crucial for forecasting realistic FPT performance at current and candidate EHT sites. The latter is achieved by automatically fetching the relevant weather quantities for the user-specified gain corruptions and noise components from the \texttt{ngehtsim} \citep{Pesce2024} database, which is derived from Modern-Era Retrospective Analysis for Research and Applications, Version 2 (MERRA-2; \citealt{Gelaro2017}) meteorological measurements. These data are retrieved when available for the observation date and when not explicitly provided by the user.\par

In the simulated observations presented here, the ideal visibilities were corrupted with only the most pertinent phase contributions: tropospheric turbulence delays and receiver and sky noise. For each site, \primes mimicks turbulence delays through a Kolmogorov-turbulent phase screen moving with a constant velocity equal to the wind speed tabulated in \texttt{ngehtsim} for the date of the observation, unless the user specifies a wind speed or a coherence time. It simulates noise using system equivalent flux densities estimated by following the procedure outlined in \citet{Simelane2024}, searching for sky brightness temperature and opacity values in the \texttt{ngehtsim} database as well if none are specified.\par

We simulated a 12-hour, dual-band observation of a point source at the position of Messier 87$^*$, starting at 00:00 UTC on 10 April 2023. The source was given flux densities of 1.1 and 1\,Jy at 86 and 230\,GHz, respectively. The \texttt{ngehtsim} weather data for the chosen date were used in order to test \pwc under realistic conditions. The phase centre was chosen to coincide with the position of the source, giving zero source structure phase. We selected an integration time of 1\,s, bandwidths of 2\,GHz centred at 86\,GHz and 230\,GHz, and an array informed by the current dual-band-capable subset of the EHT array, as well as plans for upgrades in the near-future. This includes the extended KVN (eKVN), which joined EHT campaigns in 2024 \citep{Cho2025}.  The resulting hypothetical dual-band EHT array consisted of the Atacama Pathfinder EXperiment (APEX), the Submillimeter Array (SMA), the Institut de Radioastronomie Millimétrique's (IRAM) 30-m telescope, the Greenland Telescope (GLT), the NOrthern Extended Millimeter Array (NOEMA), the Kitt Peak telescope (KP), and the KVN Pyeongchang (KVNPC) and Yonsei (KVNYS) telescopes. SMA is included assuming paired-antenna observing with SMA at the target frequency and the James Clerk Maxwell Telescope at the reference frequency \citep[e.g.,][]{Issaoun2025}. Pertinent details of the array are listed in Table~\ref{tab:eht_dualband_array}.\par

The data were calibrated using two approaches. The first was a control experiment, in which the 230-GHz data were calibrated using fringe-fitting and self-calibration (henceforth \ffsc) using the \texttt{fringefit} \citep{vanBemmel2022} and \texttt{gaincal} tasks in the Common Astronomy Software Applications (\texttt{CASA}; \citealt{McMullin2007}). We used a solution interval of 20\,s for fringe-fitting, and 10\,s for self-calibration. In the second approach, \hitops was used to apply CFPT and \pwc corrections derived from the raw simulated 86-GHz data to the 230-GHz data, saving the results from each FPT method in a separate Measurement Set (MS) column.\par

The \hitops \texttt{diagnose} mode was used to plot the phases measured on each baseline against time. \hitops was also employed to compute the coherence factor for the raw, \ffsc-calibrated, and FPT-calibrated datasets, as well as generate per-baseline comparison plots of this metric as a function of averaging interval, $T$. Finally, both calibrated datasets were imaged with \texttt{WSClean} \citep{Offringa2014} to evaluate image-plane calibration quality metrics such as the fractional peak flux recovery, the peak flux recovered in the image as a fraction of the input peak flux \citep{Rioja2023}. The peak fluxes used for this calculation were obtained by fitting 2-dimensional Gaussian profiles to the sources in the images using the \texttt{imfit} task in \texttt{CASA}.\par

The process described above, from synthetic data generation to image-plane analysis, was implemented using \texttt{Stimela2}, a system-agnostic framework for scripting scalable and reproducible workflows \citep{Smirnov2025}.

\begin{figure}
    \centering
    \includegraphics[width=\linewidth]{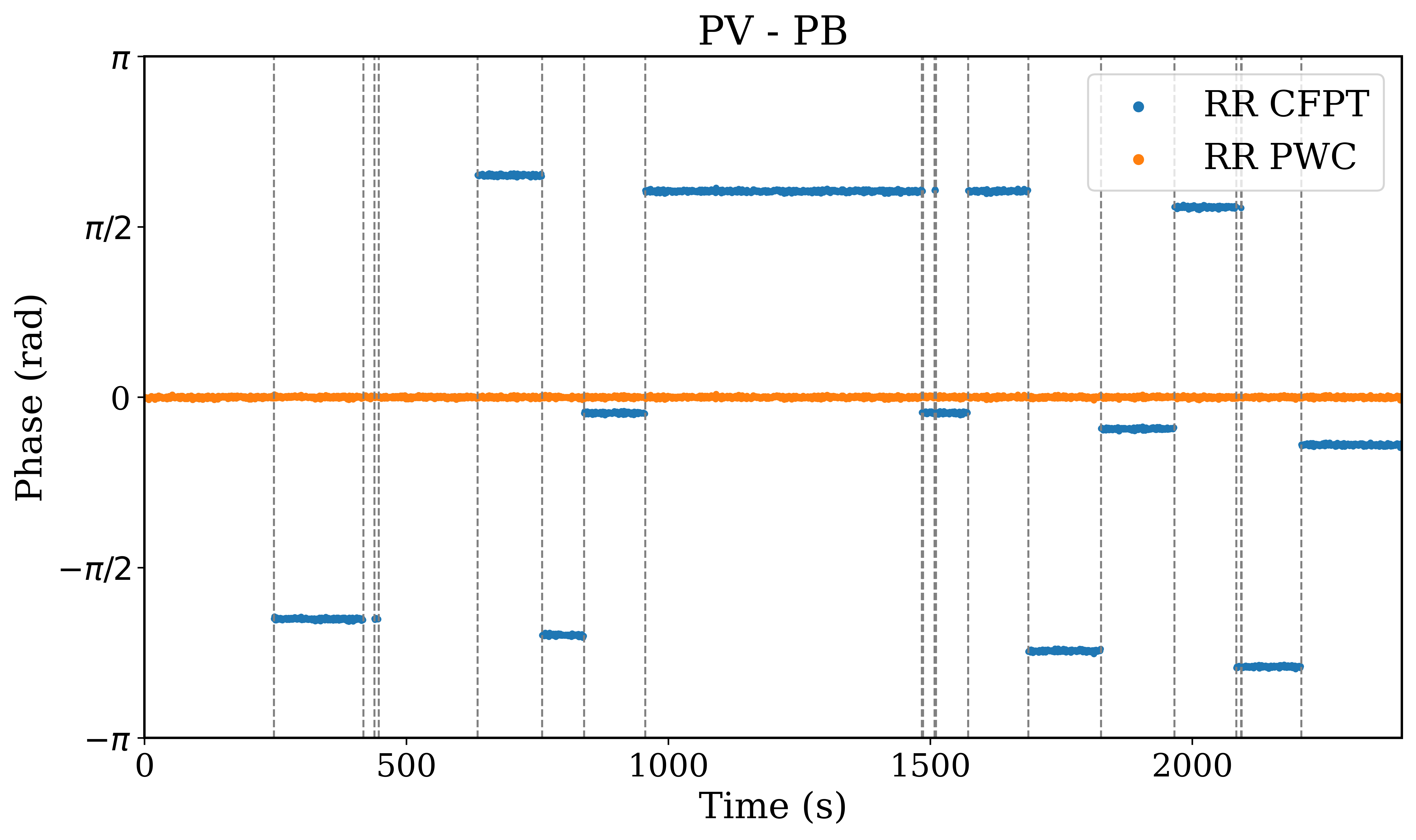}
    \caption{Residual target-frequency (230\,GHz) phase after CFPT (blue) and \pwc (orange) corrections for the baseline formed by IRAM and NOEMA for the first 40\,minutes of the observation. These are same data plotted in \fig{\ref{fig:cfpt_vs_pwc_phase}}, with vertical lines marking the times at which phase-wraps were detected in the reference-frequency (86-GHz) phases. These lines show that the jump discontinuities in the CFPT residuals coincide with phase-wraps at the reference frequency.}
    \label{fig:cfpt_vs_pwc_phase_w_wrap_positions}
\end{figure}

\section{Results and discussion} \label{sect:results_and_discussion}
Our results showed that \pwc stabilises the phases on all baselines without leaving any phase jumps like those typically observed in CFPT residuals. \fig{\ref{fig:cfpt_vs_pwc_phase}} shows an example of this using the residual phases for the baseline formed by IRAM and NOEMA, chosen for illustrative purposes as it is one of the most sensitive baselines. However, on baselines where the source was not mutually visible at the start of the observation, we observed constant phase offsets in the \pwc residuals from the expected zero-mean residuals. While the simulated data fix $n_p(t_p)=0$ at each antenna $p$ at the start of observing $t_p$, mutual visibility on a given baseline may only start after antenna $p$ has accumulated some phase-wraps, i.e., $|n_p(t_0)|>0$. In this case, the assumption that $n_0 = 0$, where $n_0 \equiv n(t_0)$ is the integer ambiguity of visibility phase $\phi(t_0)$, is incorrect. If the phase-wrap-tracking function determined by the algorithm, $n'(t)$, differs from the correct one, $n(t)$, the \pwc residuals contain phase errors given by
\begin{equation}
    \Delta \Phi_{\mathrm{PWC}}(t) = 2 \pi \epsilon [n(t)-n'(t)] \,. 
\end{equation}
Since we see only one phase offset, $\Delta \Phi_{\mathrm{PWC}}(t)$ must be a constant. Therefore, because the offset occurs from $t_0$,
\begin{align}
    \Delta \Phi_{\mathrm{PWC}}(t) &= 2 \pi \epsilon [n(t_0)-n'(t_0)] \nonumber \\
    &= 2 \pi \epsilon n_0\,,
    \label{eqn:Delta_pwc}
\end{align}
as we always set $n'(t_0)$ to zero. The discovery of these offsets informed the augmentation of the \pwc implementation in \hitops with an optional step to remove constant phase offsets using the phases of some user-provided reference data. In practice, these could be the residuals of a different calibration approach or model visibilities estimated in the data reduction. The residual phases $\Phi_\mathrm{FFSC}(t)$ served this purpose in this work. The offset removal works by computing offsets for a large range of $n$-values, $\{\Delta \Phi_i\} = 2 \pi \epsilon n_i$, as well as the mean difference between the reference residual phases and the \pwc residual phases. We use a circular mean to avoid phase-wrapping effects:
\begin{equation}
    \Delta \Phi_\mathrm{PWC} = \mathrm{Arg}(\bar{z})\,,
\end{equation}
where $\bar{z}$ is the mean of the time-series
\begin{equation}
    z(t) = \exp{ \left[ i \left(\Phi_\mathrm{FFSC}(t)-\Phi_\mathrm{PWC}(t) \right) \right] }\,.
\end{equation}
The best estimate of $n_0$ is then chosen to be the $n_i$ value whose corresponding offset is closest to $\Delta \Phi_\mathrm{PWC}$. Finally, we add the corresponding offset to $\Phi_\mathrm{PWC}(t)$.\par

These offsets (\eq{\ref{eqn:Delta_pwc}}) are likely a simulation artefact caused by artificially wrapping tropospheric turbulence-induced phase delays that were first generated as unbounded time-series, an explanation we will validate with real data. All results presented in this section were obtained with the augmented \pwc algorithm.\par

\begin{figure}
    \centering
    \includegraphics[width=\linewidth]{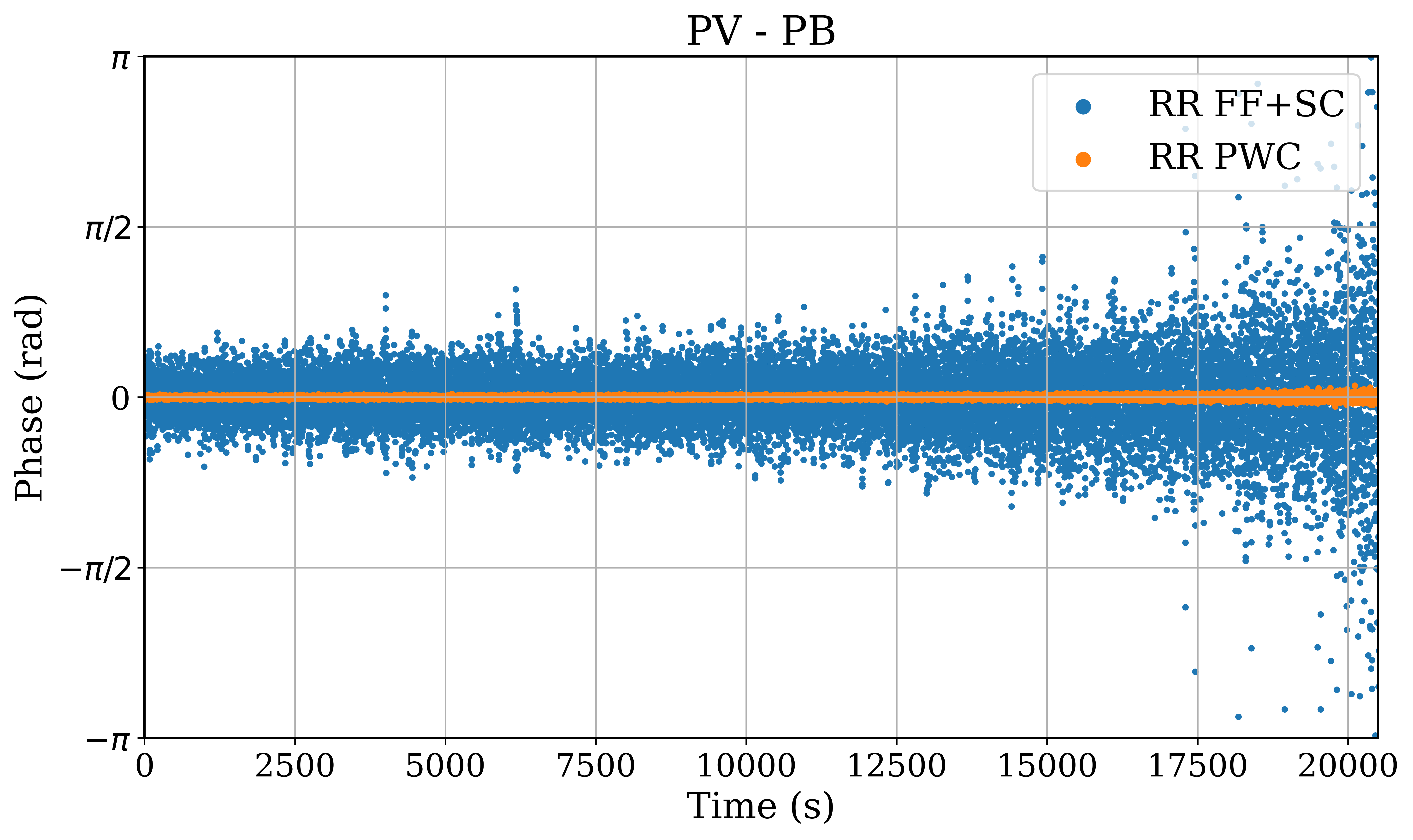}
    \caption{Residual target-frequency (230\,GHz) phase after conventional phase calibration (blue) and \pwc (orange) for the baseline formed by IRAM and NOEMA. The \pwc residuals show significantly smaller scatter.}
    \label{fig:ff_vs_pwc_phase}
\end{figure} 

The CFPT residuals show phase jumps at every phase-wrapping instance of the 86-GHz phases, marked by the vertical dashed lines in \fig{\ref{fig:cfpt_vs_pwc_phase_w_wrap_positions}}, which shows the early section of the phase data plotted in \fig{\ref{fig:cfpt_vs_pwc_phase}}. Aside from thermal phase noise, these residuals in Figures~\ref{fig:cfpt_vs_pwc_phase} and \ref{fig:cfpt_vs_pwc_phase_w_wrap_positions} provide a visualisation of the function $\delta \Phi_{\mathrm{FPT}}(t)$. As predicted by \eq{\ref{eqn:pwc_residuals}}, \pwc correctly models and subtracts this term, and its residuals are consequently free of jump discontinuities. This result suggests that, with realistic coherence times and system temperatures, accurate phase tracking is possible at 86\,GHz. At this frequency, during the nominal April observing window, current EHT stations are therefore expected to operate in an SNR regime where the Itoh condition holds, and \pwc can be expected to work, enabling this band to be used for 230-GHz non-dispersive delay correction.\par

Compared to the conventional phase calibration (\ffsc) approach employed in the control experiment, the \pwc residuals showed smaller phase scatter about the expected mean value of zero (see \fig{\ref{fig:ff_vs_pwc_phase}}). Since the thermal noise contribution to the FPT phase residuals is a linear combination of the Gaussian random variables $\phi_\mathrm{ther}(t)$ and $\Phi_\mathrm{ther}(t)$,
its variance is
\begin{equation}
    \sigma_\mathrm{FPT}^2 = \sigma_\mathrm{t}^2 + R^2 \sigma_\mathrm{r}^2\,, 
    \label{eqn:fpt_noise}
\end{equation}
where $\sigma_\mathrm{r}^2$ and $\sigma_\mathrm{t}^2$ are the variances of $\phi_\mathrm{ther}(t)$ and $\Phi_\mathrm{ther}(t)$, respectively. The relatively small scatter suggests that FPT is resilient to noise at these frequencies (86 and 230\,GHz), even with $R \approx 2.674$, despite $\sigma_\mathrm{FPT}^2$ containing a contribution from a term proportional to $R^2$. This bodes well for SNR-sensitive calibration further downstream, which might be required to mitigate residual phase errors post-FPT, such as the instrumental term in \eq{\ref{eqn:cfpt}}.\par

\begin{figure}
    \centering
    \includegraphics[width=\linewidth]{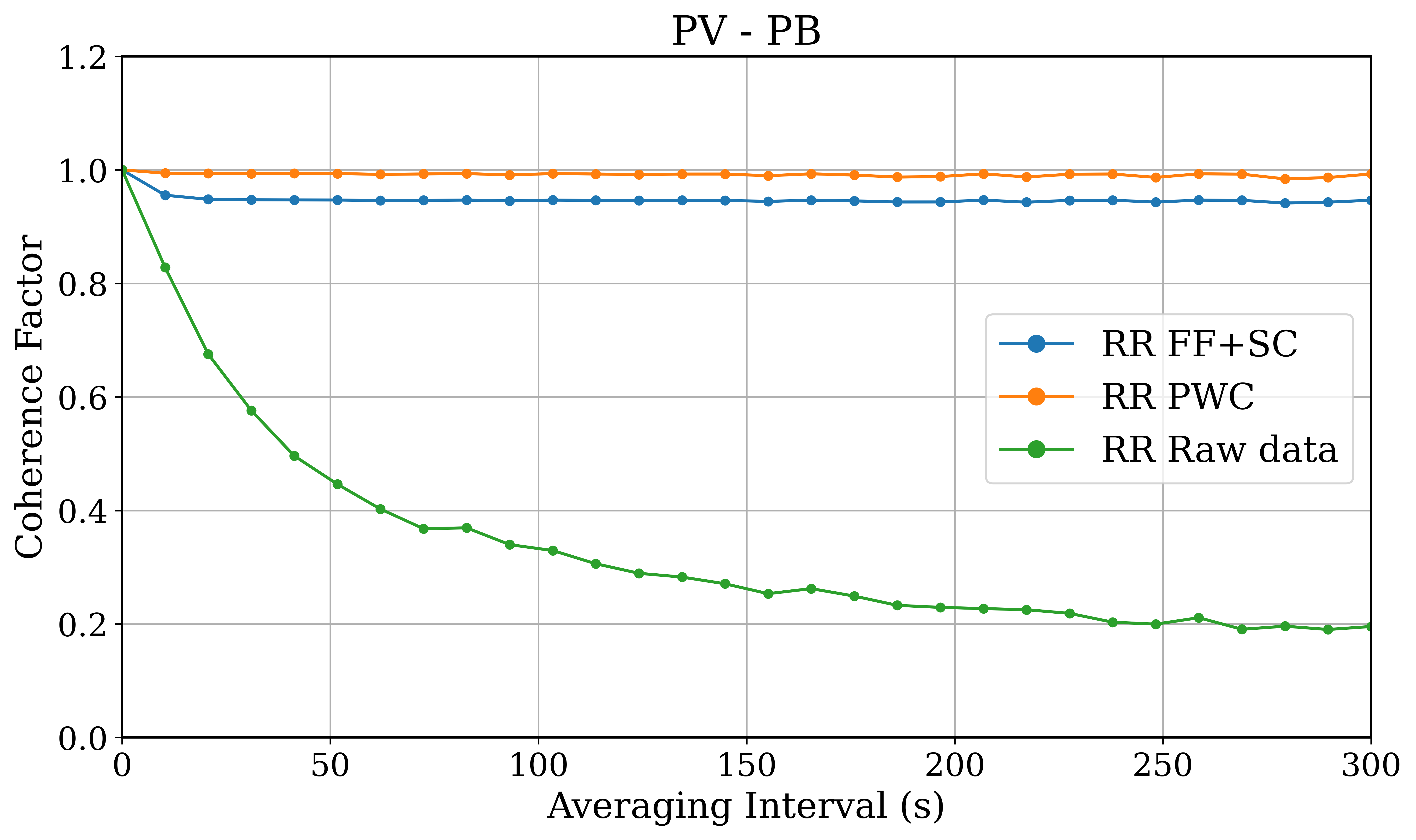}
    \caption{Coherence factor comparison between the raw data (green), conventional phase calibration (blue), and \pwc (orange) for the baseline formed by IRAM and NOEMA. The \pwc residuals show a 143 per cent coherence improvement over the raw data and a 5 per cent improvement over the \ffsc residuals.}
    \label{fig:ff_vs_pwc_coherence}
\end{figure}

\pwc achieved superior coherence compared to the reference \ffsc calibration scheme for every averaging interval on all baselines when measured through the coherence factor, $C(T)$ (see \fig{\ref{fig:ff_vs_pwc_coherence}}). In this figure, the IRAM-NOEMA (PV-PB) baseline shows a 143 per cent phase coherence improvement compared to the raw data and a 5 per cent improvement compared to the reference approach. \citet{Zhao2025} compared the post-FPT coherence factor with the coherence factor before FPT and without any delay rate removal. Since the coherence factor is sensitive only to the rates, their pre-FPT coherence factors can be compared to our raw-data coherence factors. After steep declines in the interval $0<T\lesssim100\,\mathrm{s}$, their coherence factors generally stabilised around a coherence factor of $\sim0.2$. The coherence of our raw phases exhibited broadly similar behaviour, providing a first-order validation of the approach to simulating the effect of tropospheric turbulence on millimetre signals in \primes and \texttt{MeqSilhouette}.\par

\begin{figure*}
    \centering
    \includegraphics[width=\textwidth]{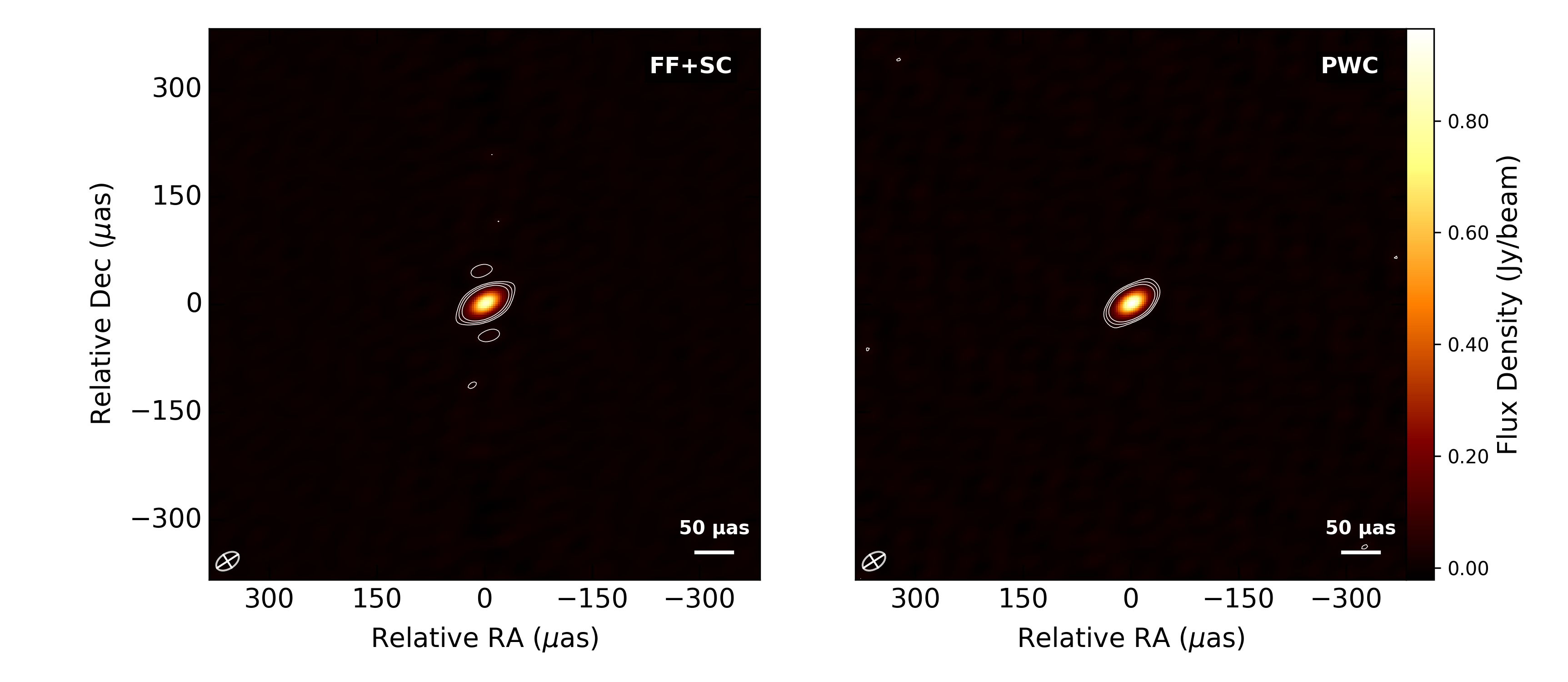}
    \caption{Comparison of 230-GHz reference image generated from \ffsc (left) and \pwc image (right). Contours (white) are plotted at the $5\sigma$, $10\sigma$, and $20\sigma$ levels, with $\sigma=3.26\,\mathrm{mJy \, beam}^{-1}$, highlighting higher sidelobe peaks in the \ffsc image. The \pwc image resembles the reference image, with a very similar dynamic range, showing that \pwc achieves a successful FPT phase calibration when R takes on a non-integer value.}
    \label{fig:images}
\end{figure*}

\begin{figure}
    \centering
    \includegraphics[width=\linewidth]{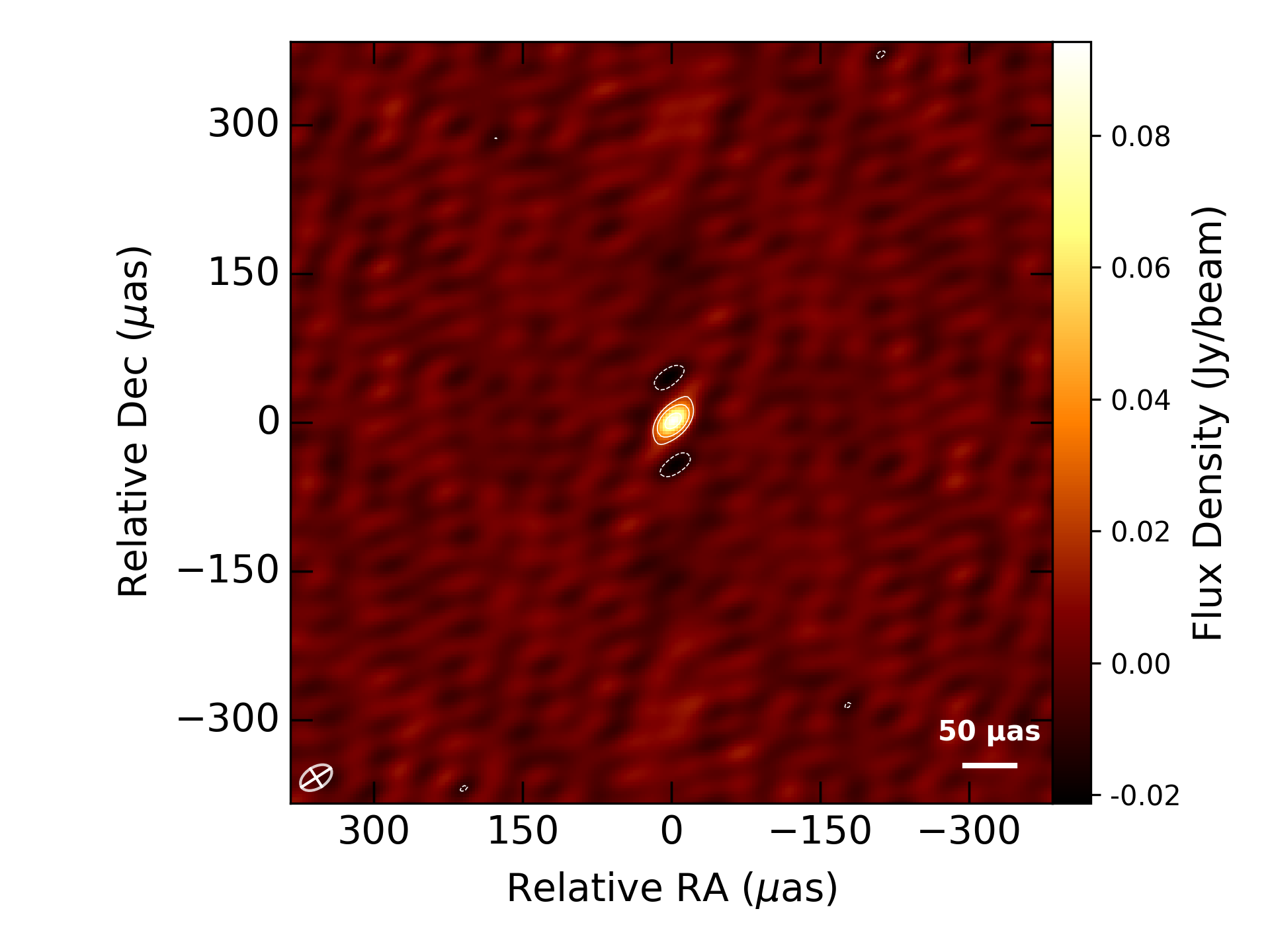}
    \caption{Difference image obtained from subtraction of the reference (\ffsc) image from the \pwc image (\fig{\ref{fig:images}}). Contours (white) are plotted at the $-3\sigma$, $5\sigma$, $10\sigma$, and $20\sigma$ levels, with $\sigma=3.67\,\mathrm{mJy \, beam}^{-1}$ and the negative contour level indicated by dashed lines. The peak flux density is $94.5\,\mathrm{mJy \, beam}^{-1}$, 9.8 per cent of the peak in the \pwc image.}
    \label{fig:difference_image}
\end{figure}

Overall, the \pwc residuals had between 3 per cent (SM-KP) and 50 per cent (AP-KC) better coherence than the \ffsc residuals. The ratio of post-FPT coherence to post-\ffsc coherence was highest in the least sensitive baselines. This is due to similar post-FPT coherence on all baselines, but declining post-\ffsc coherence with declining baseline sensitivity. The coherence of the \ffsc residuals ranged from $0.61$ (AP-KC) to $0.96$ (SM-KP), while that of the \pwc residuals ranged from $0.88$ (AP-KC) to $0.99$ (SM-KP, AP-PV, AP-PB, PV-PB, AP-GL, PV-GL, GL-PB, and GL-KP). These results suggest that the least sensitive baselines benefit more from FPT, which is promising for proposed new EHT sites with moderate weather conditions. On average, \pwc outperformed the reference approach by 18 per cent, with a median coherence improvement of 11 per cent over the reference approach. The stability of $C(T)$ with increasing $T$ in \fig{\ref{fig:ff_vs_pwc_coherence}} and similar plots for other baselines suggests that, as expected for FPT when the tropospheric delays scale perfectly with frequency, \pwc successfully removed the fringe rates on all baselines, leaving phase residuals with only noise-limited coherence.\par

\fig{\ref{fig:images}} shows a comparison of the images obtained from \ffsc and \pwc. The \pwc image resembles the reference image generated from \ffsc very well. \pwc yielded a fractional flux recovery of $0.965 \pm 0.0046$, while the reference approach only achieved $0.868 \pm 0.0045$, representing relative errors of approximately 4 and 13 per cent, respectively. This suggests that the constant phase offset removal step with which the \pwc algorithm proposed in Section~\ref{sect:methods} was supplemented finds the correct $n_0$ values. The difference image (\pwc image minus reference image) is shown in \fig{\ref{fig:difference_image}}. It shows \pwc recovering a peak flux that is $94.5 \pm 3.4\,\mathrm{mJy \, beam}^{-1}$ higher than \ffsc, further strengthening the argument that the \pwc residual phases are likely correct. Additionally, regions of negative pixels can be seen at the peaks of the sidelobes of the point spread function, possibly suggesting a higher-quality deconvolution in the imaging of \pwc-calibrated data.\par

For this work, processing was performed on a server equipped with Intel Xeon Silver 4216 CPUs (32 hyperthreaded cores clocked at 2.1\,GHz). Calibration of the 230-GHz dataset using PWC took less than 20\,s, requiring only two logical cores (effectively a single physical core) and 6.5\,GB of memory. This demonstrates that the \hitops implementation of PWC is computationally feasible for application to EHT datasets.\par

Based on the results of the simple simulations presented in this work, \pwc appears viable for EHT frequencies, but tests including more challenging features of real data are necessary to further validate and develop the \pwc calibration algorithm. For example, challenges yet to be tackled include intra-scan time-gaps due to flagged data and large ($\geq \pi$) instrumental phase jumps. \primes also provides all of the functionality required to empirically translate the Itoh condition into a noise limit up to which the rudimentary phase-tracking approach of the current \pwc implementation will succeed. This will be useful as EHT expansion projects look towards geographically advantageous candidate sites that may have sub-optimal weather conditions for millimetre observing \citep[see the candidate site surveys by][for example]{Raymond2021, Pesce2024, Simelane2024, Frans2025}. Additionally, as the ngEHT project aims for year-round observing \citep{Doeleman2023}, such a limit would help determine the periods in the year during which the approach will work for existing and candidate EHT sites. Lastly, since the principal science targets of the EHT are spatially resolved black hole shadows, it will be important to test the resilience of FPT against errors in source structure phase estimation at the reference frequency, particularly for rapidly evolving sources such as Sagittarius A$^*$.

\section{Summary and conclusions} \label{sect:conclusions}
We have illuminated the nature of the jump discontinuities in residual phases reported by previous works \citep[e.g.,][]{Dodson2014, Rioja2023, Issaoun2025} when performing FPT with non-integer frequency ratios. The algebra presented in this paper shows that the phase jumps are, in fact, deterministic and correctly predicts their values.\par

Using this knowledge, we developed an FPT approach that works regardless of the frequency ratio, provided that the phases recorded at the reference frequency satisfy the Itoh condition. This algorithm, called phase-wrap counting (\pwc), has been implemented in a Python package called \hitops. Classical FPT approaches are also available through the package, alongside tools for analysis and visualisation of FPT performance through various coherence metrics, including the coherence time and coherence factor.\par

Through a realistic simulated interferometric observation generated with another new software tool (\primes), we have shown that \pwc can work at EHT frequencies, despite relying on accurate phase-wrap tracking, which previous works recommended avoiding, citing difficulty in the low-SNR regime. In contrast to those of the classical FPT approach, the \pwc phase residuals obtained in this experiment are free of jump discontinuities caused by phase-wraps at the reference frequency. They also exhibit enhanced phase coherence over the traditional phase calibration approach (fringe-fitting and self-calibration). Successful FPT, without introducing jump discontinuities, was demonstrated for image reconstruction from high-frequency data with a non-integer frequency ratio for the first time, albeit simulated. The FPT image achieved improved peak flux recovery over that obtained through the conventional data reduction approach.\par

\pwc shows significant promise for future EHT observations at 86 and 230\,GHz, but it has only been tested on simulated data so far. The added nuances that are not captured by simulated data necessitate testing the approach on real observations.

\section*{Acknowledgements}
The authors thank Geoff Beck, Coral Pillay, Mika Naidoo, Lindy Blackburn, Dominic Pesce, and Iniyan Natarajan for insightful discussions during the course of this work. The financial assistance of the South African Radio Astronomy Observatory (SARAO) towards this research is hereby acknowledged (www.sarao.ac.za). RPD’s research is funded by the South African Research Chairs Initiative of the DSTI/NRF. This work made use of the CARTA (Cube Analysis and Rendering Tool for Astronomy) software (DOI: 10.5281/zenodo.3377984 – https://cartavis.github.io).

\section*{Data Availability}
Simulation data may be made available upon reasonable request.



\bibliographystyle{rasti}
\bibliography{references} 








\bsp	
\label{lastpage}
\end{document}